\documentclass[hyper,11pt,letterpaper]{JHEP3}
\usepackage{graphicx, subfigure, amsmath,amssymb,calrsfs,mathrsfs,url}

\newcommand{\tr}{\text{tr }}

\newcommand{\mc}[1]{\mathscr{#1}}
\newcommand{\rv}[1]{\boldsymbol{\mathsf{#1}}}

\newcommand{\vev}[1]{\left\langle #1\right\rangle}

\title{Euclidean Quantum Mechanics and Universal Nonlinear Filtering}
\author{Bhashyam Balaji\\
Radar Systems Section,\\
Defence Research and Development Canada, Ottawa,\\
3701 Carling Avenue, \\
Ottawa ON K1A 0Z4 Canada\\
Email: Bhashyam.Balaji@drdc-rddc.gc.ca\\
}

\maketitle

\abstract{
	An important problem in applied science is the continuous nonlinear filtering problem, i.e., the estimation of a Langevin state that is observed indirectly. In this paper, it is shown that Euclidean quantum mechanics is closely related to the continuous nonlinear filtering problem. The key is the configuration space Feynman path integral representation of the fundamental solution of a Fokker-Planck type of equation termed the Yau Equation of continuous-continuous filtering. A corollary is the equivalence between nonlinear filtering problem and a time-varying Schr\"odinger equation.
}

\keywords{
	Fokker-Planck Equation, Kolmogorov Equation, stochastic differential equations, Duncan-Mortensen-Zakai (DMZ) Equation, nonlinear filtering, Feynman path integral
}
\begin{document}

\section{Introduction}
 The problem of the evolution of a Langevin state, or a signal of interest, described by a continuous-time stochastic dynamical model arises frequently in practice. Specifically, in (classical) filtering probelms one deals with macroscopic objects whose state variables  are phenomenologically well-described by classical deterministic laws modified by external disturbances that can be modelled as random noise. So the state of the system  is described by a noisy version of a deterministic nonlinear dynamical system termed the state model, i.e., the dynamics is governed by a system of first-order differential equations in the state variable ($\rv{x}(t)$) with an additional contribution due to noise that is random ($\rv{\nu}(t)$).  The noise in the state model is referred to as the signal noise. If the noise is Gaussian (or more generally  multiplicative Gaussian) the state process is a Markov process. Since the process is stochastic, the state process is completely characterized by a probability density function. The Fokker-Planck-Kolmogorov foward equation (FPKfe) describes the evolution of this probability density function (or equivalently, the transition probability density function) and is the complete solution of the state evolution problem\cite{H.Risken1999}.

 However, in many applications the signal, or state variables, cannot  be directly observed. Instead, what is measured is a related stochastic process ($\rv{y}(t)$) called the measurement process. The measurement process can often be modelled as yet another nonlinear continuous-time stochastic dynamical system called the measurement model. Loosely speaking, the observations, or measurements, are discrete-time samples drawn from a different system of noisy first order differential equations. The noise in the measurement dynamical system is referred to as measurement noise. The continuous-continuous ``filtering'' problem is the estimation of the continuous-time signal or state process given the (noisy) observations of a related continuous-time stochastic process (the measurement or observation process). For a recent discussion, see \cite{Balaji2008}.

The conditional probability density is the complete solution of the filtering problem. That is, it embodies all the probabilistic information about the state process contained in the observations and in the initial condition. Here the Bayesian point of view is being adopted.  From the conditional probability density, an optimal estimate may be computed for any loss function. For instance, the minimum variance estimate is the conditional mean. The solution of the nonlinear filtering problem is said to be universal if the initial distribution can be arbitrary\footnote{A classic discussion of nonlinear filtering theory can be found in the text by Jazwinski\cite{A.H.Jazwinski1970}. For a more up-to-date discussion, see \cite{PAPER2}..}. 


The traditional approach to the solution of the continuous-continuous universal nonlinear filtering problem requires the solution of the Duncan-Mortensen-Zakai (DMZ) equation, a stochastic differential equation  (SDE) describing the unnormalized conditional probability density. The DMZ equation can be gauge transformed to the time-varying partial differential equation termed the robust DMZ equation. However, since the robust DMZ equation is a partial differential equation (PDE) with the coefficients depending on the measurements the PDE cannot be solved off-line, or in real time. In \cite{S.T.YauS.-TYau2000},  it was proved that solving the robust DMZ equation is equivalent to solving a PDE, termed the Yau Equation (YYe) of continuous-continuous filtering, whose coefficients are independent of the measurements. Hence, the YYe can be solved off-line and also in a memoryless way. 

Recently, it has been noted that the Feynman path integral\footnote{For a classic review of path integrals, see \cite{R.P.FeynmanandA.R.Hibbs1965}.} can be used to solve the general nonlinear filtering problem. In fact, it was shown that the path integral formulation of the continuous-continuous filtering problem directly leads to the YYe\cite{PAPER2}. The path integral formula for the fundamental solution of the YYe was also derived in \cite{PAPER2}. The advantages of the Feynman path integral formulation  are many, including a completely independent and self-contained formulation and solution of the general nonlinear filtering problem (than is traditional in filtering theory literature which is based on measure theoretic techniques termed the Feynman-K\v ac formalism), and simple, efficient and accurate algorithms implementable in real-time\cite{Balaji2008}.

The Feynman path integral has proven to be a very powerful tool in modern theoretical physics, often leading to results that are not evident using other methods. In this paper, one such instance is provided in the application to the filtering problem. In particular, it is demonstrated that there is a very close relationship between nonlinear filtering theory and Euclidean quantum mechanics. Specifically, the fundamental solution of the YYe may be viewed as the expectation of a certain operator in a Euclidean quantum mechanical system. This follows from the path integral formula for the fundamental solution of the YYe derived in \cite{PAPER2}. This result generalizes the equivalence between nonlinear filtering and the Euclidean Schr\"odinger equation derived  in \cite{YauS-T.Yau2004} via more complicated means. Their result was limited to a filtering system where the signal model drift is the same as in the finite-dimensional Yau filter\footnote{Their explicit solution imposed an additional condition on the measurement model drift, but that is not relevant for our discussion here.}. This result does not require that assumption---the signal model is assumed to have additive noise (or somewhat more generally, orthogonal diffusion vielbein so that the diffusion matrix is proportional to the identity matrix), and the measurement model noise is also additive.  Also, no explicit time dependence is assumed in the model. 

The outline of our paper is as follows. In Section \ref{sec:BasicResults}, notation and some of the important results for the real-time solution of the continuous-continuous nonlinear filtering problem are reviewed. In Section \ref{sec:PIFormulaYau}, are summaries of the results obtained using the Feynman path integral formulation, specifically the path integral expressions for the fundamental solution for the FPKfe and the YYe. In Section \ref{sec:YEandEucQM}, the equivalence between Euclidean quantum mechanics and nonlinear filtering is presented. In the following section it is shown that it leads to the equivalence between nonlinear filtering and the time-varying Schr\"odinger equation obtained in \cite{YauS-T.Yau2004}. In Section \ref{sec:AdditionalRemarks}, some important conceptual issues are discussed which further clarify the relationship between the YYe and Euclidean quantum mechanics. The conclusions and directions of future work are presented in Section \ref{sec:Conclusion}. In Appendix \ref{sec:VerifyYau}, the path integral formula for the fundamental solution of the YYe that forms the basis of our work is verified.

\section{Basic Results of Nonlinear Filtering} \label{sec:BasicResults}
A very brief summary of results of filtering theory is provided in this section. For the purposes of this paper, the important result is that the solution of the filtering problem requires the solution of a Fokker-Planck type of equation called the Yau equation. A more complete discussion with references can be found in \cite{PAPER2} and \cite{Balaji2008}. It is worth contrasting the simplicity of the Feynman path integral solution discussed in the next section with the traditional discussion summarized here.  

\subsection{From the DMZ SDE to the Robust DMZ PDE}
The signal and observation model considered is the following:
\begin{align}
  \label{eq:signalobse}
  \begin{cases}
  d\rv{x}(t)&=f(\rv{x}(t))dt+e(\rv{x}(t))d\rv{v}(t),\quad x(0)=x_0,\\ 
  d\rv{y}(t)&=h(\rv{x}(t))dt+d\rv{w}(t),\quad y(0)=0.
  \end{cases}
\end{align}
Here $\rv{x}$ and $\rv{v}$ are $\mathbb{R}^n-$valued stochastic processes, and $\rv{y}$ and $\rv{w}$ are $\mathbb{R}^m-$valued stochastic processes, respectively, and $e\in\mathbb{R}^{n\times n}$. These are defined in the It\^o sense. The processes $\rv{v}$ and $\rv{w}$ are assumed to be independent Brownian processes with variances $\hbar_{\nu}$ and $\hbar_{\mu}$ respectively.  Also, $f(h)$ is referred to as the signal (measurement) model drift, $e$ as the diffusion vielbein, and $ee^T$ as the diffusion matrix. In this paper, the additive noise case is considered where $e$ is the identity matrix, although the same result holds if one assumes that the diffusion vielbein is orthogonal. Finally, no explicit time dependence is assumed in the model.

The unnormalized conditional probability density, $\sigma(t,x)$, of the state given the observations $\{Y(s):0\leq s\leq t\}$ satisfies the DMZ stochastic differential equation:
\begin{align}
  \label{eq:dmzsde}
  d\sigma(t,x)=\mc{L}_Y\sigma(t,x)dt+\sum_{i=1}^m\mc{L}_i\sigma(t,x)dy_i(t),\quad\text{where } \sigma(0,x)=\sigma_0(x).
\end{align}
Here 
\begin{align}
  \label{eq:L0}
  \mc{L}_Y(\sigma(t,x))=-\sum_{i=1}^n\frac{\partial}{\partial x_i}\left( f_i(x)\sigma(t,x) \right)+\frac{1}{2}\sum_{i=1}^n\frac{\partial^2\sigma}{\partial x_i^2}(t,x)-\frac{1}{2}\sum_{i=1}^mh_i^2(x)\sigma(t,x),\\ \nonumber
\end{align}

where $\mc{L}_i$ is the zero-degree differential operator $h_i(x), i=1,\ldots,m$, $\sigma_0(x)$ is the probability density at the initial time $t_0=0$. Under the following gauge transformation 
\begin{align}
  \label{eq:davistransf}
  u(t,x)=\exp\left( -\sum_{i=1}^mh_i(x)y_i(t) \right)\sigma(t,x),
\end{align}
the DMZ SDE is transformed into the following time-varying PDE called the robust DMZ equation \cite{S.-T.YauS.S.-TYau1996}:
\begin{align}
\label{eq:pdedmz}
\frac{\partial u}{\partial t}(t,x)=&\frac{1}{2}\sum_{i=1}^n\frac{\partial^2 u}{\partial x_i^2}(t,x)
+\sum_{i=1}^n\left( -f_i(x)+\sum_{j=1}^my_j(t)\frac{\partial h_j}{\partial x_i}(x) \right)\frac{\partial u}{\partial x_i}(t,x)\\\nonumber
&-\biggl( \sum_{i=1}^n\frac{\partial f_i}{\partial x_i}(x)+\frac{1}{2}\sum_{i=1}^mh_i^2(x)-\frac{1}{2}\sum_{i=1}^my_i(t)\Delta h_i(x)+\sum_{i=1}^m\sum_{j=1}^ny_i(t)f_j(x)\frac{\partial h_i}{\partial x_j}(x)\\ \nonumber
&\quad-\frac{1}{2}\sum_{i,j=1}^m\sum_{k=1}^ny_i(t)y_j(t)\frac{\partial h_i}{\partial x_k}(x)\frac{\partial h_j}{\partial x_k}(x) \biggr)u(t,x),\\  \nonumber
u(0,x)=&\sigma_0(x).
\end{align}
Here $\Delta$ is the Laplacian. The solution of a PDE when the initial condition is a delta function is called its fundamental solution.

\subsection{The Yau Equation of Continuous-Continuous Filtering}

Recently, S-T. Yau and Stephen Yau made a major advance in the real-time solution of the general nonlinear filtering problem \cite{S.T.YauS.-TYau2000}. They began by observing that if $u_l(t,x)$ satisfies the equation 
	\begin{align}
\label{eq:pdedmzfrozen}
\frac{\partial u_l}{\partial t}(t,x)=&\frac{1}{2}\sum_{i=1}^n\frac{\partial^2 u_l}{\partial x_i^2}(t,x)
+\sum_{i=1}^n\left( -f_i(x)+\sum_{j=1}^my_j(\tau_l)\frac{\partial h_j}{\partial x_i}(x) \right)\frac{\partial u_l}{\partial x_i}(t,x)\\\nonumber
&-\biggl( \sum_{i=1}^n\frac{\partial f_i}{\partial x_i}(x)+\frac{1}{2}\sum_{i=1}^mh_i^2(x)-\frac{1}{2}\sum_{i=1}^my_i(\tau_l)\Delta h_i(x)+\sum_{i=1}^m\sum_{j=1}^ny_i(\tau_l)f_j(x)\frac{\partial h_i}{\partial x_j}(x)\\ \nonumber
&\quad-\frac{1}{2}\sum_{i,j=1}^m\sum_{k=1}^ny_i(\tau_l)y_j(\tau_l)\frac{\partial h_i}{\partial x_k}(x)\frac{\partial h_j}{\partial x_k}(x) \biggr)u_l(t,x),\\  \nonumber
u_l(\tau_{l-1},x)=&u_{l-1}(\tau_{l-1},x),
\end{align}
in the time interval $\tau_{l-1}\le t\le\tau_l$, then the function $\tilde{u}_l(t,x)$ defined as
\begin{align}\label{eq:postveryau}
	\tilde{u}_l(t,x)=\exp\left( \sum_{i=1}^my_i(\tau_l)h_i(x) \right)u_l(t,x)
\end{align}
satisfies the parabolic partial differential equation termed the Yau Equation (YYe)
\begin{align}
	\label{eq:yauONFeqn}
	\frac{\partial \tilde{u}_l}{\partial t}(t,x)=\frac{1}{2}\sum_{i=1}^n\frac{\partial^2\tilde{u}_l}{\partial x_i^2}(t,x)-\sum_{i=1}^nf_l(x)\frac{\partial \tilde{u}_l}{\partial x_i}(t,x)-\left( \sum_{i=1}^n\frac{\partial f_i}{\partial x_i}(x)+\frac{1}{2}\sum_{i=1}^mh_i^2(x) \right)\tilde{u}_l(t,x)
\end{align}
in the same time interval. The converse of the statement is also true. In \cite{S.S.-T.YauandS.-T.Yau2001}, they further showed that it is sufficient to use the previous observation, i.e., $u_l(t,x)$ satisfies Equation \ref{eq:pdedmzfrozen} if and only if $\tilde{u}_l(t,x)$ defined as
\begin{align}\label{eq:preveryau}
	\tilde{u}_l(t,x)=\exp\left( \sum_{i=1}^my_i(\tau_{l-1})h_i(x) \right)u_l(t,x)
\end{align}
satisfies Equation \ref{eq:yauONFeqn} in the time interval $\tau_{l-1}\le t\le\tau_l$.  Equation \ref{eq:postveryau} ( Equation \ref{eq:preveryau}) and Equation \ref{eq:yauONFeqn} are referred to as the post-measurement (pre-measurement) forms of the YYe.  

Equation \ref{eq:pdedmzfrozen} can also be obtained by setting $y(t)$ to $y(\tau_l)$ in Equation \ref{eq:pdedmz}. It was proved that the solution of Equation \ref{eq:pdedmzfrozen} approximates very well the solution of the robust DMZ equation (Equation \ref{eq:pdedmz}), i.e., it converges to $u(t,x)$ in both the pointwise sense and the $L^2$ sense. Thus, solving Equation \ref{eq:pdedmz} is equivalent to solving Equation \ref{eq:yauONFeqn}. In fact, in \cite{YauYau2005} it was proved that $\tilde{u}(t,x)$ converges to $\sigma(t,x)$.

There are two important points to note. The coefficients of the robust DMZ equation (Equation \ref{eq:pdedmz}) contain the measurements. Consequently, the partial differential equation to be solved for continuous-continuous filtering is unknown prior to measurements. In other words, the robust DMZ equation has to be solved on-line; its solution cannot be pre-computed. By contrast, in the YYe (Equation \ref{eq:yauONFeqn}), measurements are absent from the partial differential equation. The measurements only enter the initial condition at each measurement step. This implies that the YYe can be be solved  off-line; there is no need for on-line solution of PDEs\footnote{This assumes that the measurement steps are equidistant or known.}. This makes real-time solution feasible even if the state dimension is large. The algorithm based on the YYe is termed the Yau algorithm

In addition, it is noted that all other known algorithms for continuous-continuous filtering assume boundedness of the measurement model drift (see discussion in \cite{YauYau2005} and \cite{YauYau2008}). This is a highly restrictive assumption and implies, for instance, that they cannot even handle the linear model which the Kalman filter handles very well. The Yau algorithm is shown to converge to the true solution under much weaker conditions on the drift (see \cite{S.T.YauS.-TYau2000},\cite{S.S.-T.YauandS.-T.Yau2001}, \cite{YauYau2008}). In that sense, the Yau algorithm is the \textit{only practical nonlinear filtering algorithm for the general filtering problem}. Also note that from a path integral perspective, the Yau algorithm is the most natural one. 

The simplicity of the YYe has already led to some important advances. In particular, it has led to a demonstration of an equivalence between solving a class of nonlinear filtering problems and the time-varying Schr\"odinger equation\cite{YauS-T.Yau2004}. As a result, the Yau PDE can be reduced to a system of ODEs explicitly solvable  using the power series method. This result will be shown to be a corollary of the main result of this paper. 

\section{Path Integral Formula for the Fundamental solution of the YYe}\label{sec:PIFormulaYau}
In recent papers, it was demonstrated that the path integral method leads to an independent formulation and solution of the universal nonlinear filtering problem \cite{PAPER1, PAPER2}. Specifically, the path integral formula for the transition probability density in continuous-discrete and continuous-continuous filtering was investigated. One of the results of the aforementioned work is the path integral formula for the fundamental solution of the FPKfe and the YYe. Some of the relevant results from the papers are reviewed below.

When the diffusion matrix is $\hbar_{\nu}I$, the transition probability satisfies the FPKfe (see, for instance, \cite{H.Risken1999}) 
\begin{align}
	\frac{\partial p}{\partial t}(t,x)&=-\sum_{i=1}^n\frac{\partial }{\partial x_i}\left[f_i(x(t),t)p(t,x)\right]+\frac{\hbar_{\nu}}{2}\sum_{i=1}^n\frac{\partial^2p }{\partial x_i^2} (t,x) ,\\ \nonumber
	&\equiv{\mc{L}}p(t,x).
\end{align}

The fundamental solution of this FPKfe can be shown to be given by (see \cite{JeanZinn-Justin2002} and \cite{PAPER1})
\begin{align}\label{eq:PIformula}
	P(t,x|t_0,x_0)=\int_{x(t_0)=x_0}^{x(t)=x}[\mc{D}x(t)]\exp\left( -\frac{1}{2\hbar_{\nu}}\sum_{i=1}^n\int_{t_{0}}^t dt\left[\left( \dot{x}_i(t)-f_i(x(t)) \right)^2 +\hbar_{\nu}\frac{\partial f_i}{\partial x_i}(x(t))\right]\right).
\end{align}

In continuous-continuous filtering, it is necessary to incorporate the measurement stochastic process as well as the signal process. That is, the measurement process ensemble must also be considered. The inclusion of measurement noise means that each system in the ensemble leads to a different time-dependent vector $y(t)$. Although  only one realization of the measurement stochastic process is observed, it is still meaningful to talk about an ensemble average of the measurement process (in addition to an ensemble average over the state process). Thus, the quantity of interest in continuous-continuous filtering is 
\begin{align}
	P(t_ix_i;y(t_i)|t_{i-1},x_{i-1};y(t_{i-1}))=\vev{\vev{\delta^n(\rv{x}(t_i)-x_i)\delta^m(\rv{y}(t_i)-y(t_i))}_{\rv{\mu}}}|_{\rv{x}(t_{i-1})=x_{i-1},\rv{y}(t_{i-1}=y_{i-1})},
\end{align}
where $\vev{\cdot}_{\rv{\mu}}$ denotes averaging with respect to measurement noise $\rv{\mu}(t)$. It has been shown in \cite{PAPER2} that the transition probability density conditional on the measurements is given by\footnote{Here the ``post-measurement'' form is used; the pre-measurement form does not affect the conclusion\cite{PAPER2}.}
\begin{align}\label{eq:PITimeIndepSamp}
	P(t_i,x_i;y(t_i)|t_{i-1},x_{i-1};y(t_{i-1}))
	=\exp\left(\frac{1}{\hbar_{\mu}} \sum_{k=1}^mh_k(x(t_i))\left[y_k(t_i)-y_k(t_{i-1})\right] \right)\tilde{P}(t_i,x_i|t_{i-1},x_{i-1}) 
\end{align}
where 
\begin{align}\label{eq:Yaukernel}
	\tilde{P}(t_i,x_i|t_{i-1},x_{i-1}) =\int_{x(t_{i-1})=x_{i-1}}^{x(t_i)=x_i}[\mc{D}x(t)]\exp\left(-\frac{1}{\hbar_{\nu}}S(t_{i-1},t_i)\right),
\end{align}
and
\begin{align}
	S(t_{i-1},t_i)= \frac{1}{2}\int_{t_{i-1}}^{t_i}dt\left[\sum_{i=1}^n(\dot{x}_i(t)-f_i(x(t)))^2+\hbar_{\nu}\sum_{i=1}^n\frac{\partial f_i}{\partial x_i}(x(t)) +\frac{\hbar_{\nu}}{\hbar_{\mu}}\sum_{k=1}^mh_k^2(x(t))\right]. 
\end{align}
It can be shown that $\tilde{P}(t_i,x_i,t_{i-x},x_{i-1})$ is the fundamental solution of the YYe\cite{PAPER2}. For completeness, a verification of this result is presented in the Appendix.

The transition probability density is the complete solution to the continuous filtering problem. Thus if the initial distribution is $u(t_0,x)$, then the evolved conditional probability distribution is given by
\begin{align}
	u(t,x)=\int P(t,x;y(t)|t_0,x_0';y(t_0))u(t_0,x_0)\left\{ d^nx_0 \right\}.
\end{align}

\section{YYe and Euclidean Quantum Mechanics }\label{sec:YEandEucQM}

In this section it is shown that there is a general equivalence between the YYe and Euclidean quantum mechanics. 

\subsection{A General Equivalence}\label{ssec:AGenEquiv}

Consider the path integral formula for the fundamental solution  of the Yau Equation
\begin{align}
	\tilde{P}(t,x|t_0,x_0)=\int_{x(t_0)=x_0}^{x(t)=x}[\mc{D}x(t)]\exp\left( -\frac{1}{\hbar_{\nu}}S \right),
\end{align}
where the action $S$ is
\begin{align}
	S=\frac{1}{2}\int_{t_0}^tdt\left[ \sum_{k=1}^m\left[ \dot{x}_i^2(t)+f^2_i(x(t))-2\dot{x}_i(t)f_i(x(t)) \right]+\hbar_{\nu}\sum_{i=1}^n\frac{\partial f_i}{\partial x_i}(x(t))+\frac{\hbar_{\nu}}{\hbar_{\mu}}\sum_{k=1}^mh_k^2(x(t)) \right].
\end{align}
Integrating by parts, the third term above yields
\begin{align}
	-\sum_{i=1}^n\int_{t_0}^tdt\dot{x}_i(t)f_i(x(t))&=-\sum_{i=1}^n\int_{x(t_0)}^{x(t)}dx_i(t)f_i(x(t)).
\end{align}
Here the Feynman convention is used implicitly, i.e., symmetrized arguments of $f_i$, so that the ordinary rules of calculus apply.  
Thus, the action simplifies to 
\begin{align}
	S=\frac{1}{2}\int_{t_0}^{t}dt\left[ \sum_{i=1}^n\left[ \dot{x}_i^2(t)+f_i^2(x(t)) \right]+\hbar_{\nu}\sum_{i=1}^n\frac{\partial f_i}{\partial x_i}(x(t))+\frac{\hbar_{\nu}}{\hbar_{\mu}}\sum_{k=1}^mh_k^2(x(t)) \right]-\sum_{i=1}^n\int_{x(t_0)}^{x(t)}dx_i(t)f_i(x(t)).
\end{align}
We can therefore write the fundamental solution as 
\begin{align}
	\int_{x(t_0)=x_0}^{x(t)=x}[\mc{D}x(t)]\exp\left( \frac{1}{\hbar_{\nu}}\sum_{i=1}^n\int_{x(t_0)=x_0}^{x(t)=x}dx_i(t)f_i(x(t)) \right)\exp\left(-\frac{1}{\hbar_{\nu}}\int_{t_0}^tdt\mc{L}\right),
\end{align}
where the ``Lagrangian'' $\mc{L}$, is
\begin{align}\label{eq:LagEquiv}
	\mc{L}&=\frac{1}{2}\sum_{i=1}^n\dot{x}_i^2(t)+\frac{1}{2}\sum_{i=1}^n\left[f_i^2(x(t)) +\hbar_{\nu} \frac{\partial f_i}{\partial x_i}(x(t))\right]+\frac{\hbar_{\nu}}{2\hbar_{\mu}}\sum_{k=1}^mh_k^2(x(t)),\\ \nonumber
&=T-V,
\end{align}
and  
\begin{align}
T&=\frac{1}{2}\sum_{i=1}^n\dot{x}_i^2(t),\\ \nonumber
	-V&=\frac{1}{2}\sum_{i=1}^n\left[f_i^2(x(t))+\hbar_{\nu}\frac{\partial f_i}{\partial x_i}(x(t))\right]+\frac{\hbar_{\nu}}{2\hbar_{\mu}}\sum_{k=1}^mh_k^2(x(t)).
\end{align}

Consider a Euclidean quantum mechanical system with the Hamiltonian
\begin{align}\label{eq:EqHamiltonian}
	\mc{H}=T+V. 
\end{align} 
Then, performing the path integral quantization of this system  will lead to the following expression of the transition probability amplitude:
\begin{align}
	\vev{x,t|x_0,t_0}=\int_{x(t_0)=x_0}^{x(t)=x}[\mc{D}x(t)]\exp\left(-\frac{1}{\hbar_{\nu}}\int_{t_0}^tdt\mc{L}\right).
\end{align}
Now, let us consider the following matrix element:
\begin{align}
	\vev{x,t\Big|T\left[\exp\left( \frac{1}{\hbar_{\nu}}\sum_{i=1}^n\int_{x(t_0)=x_0}^{x(t)=x}d\hat{x}_i(t)f_i(\hat{x}(t)) \right)\right]|x_0,t_0}	= \\ \nonumber
\int_{x(t_0)=x_0}^{x(t)=x}[\mc{D}x(t)]\exp\left( \frac{1}{\hbar_{\nu}}\sum_{i=1}^n\int_{x(t_0)=x_0}^{x(t)=x}dx_i(t)f_i(x(t)) \right)\exp\left(-\frac{1}{\hbar_{\nu}}\int_{t_0}^tdt\mc{L}\right).
\end{align}

In the general case, the value of the line integral depends on the path. Therefore, it cannot be ``factored out''. After all, it is a part of the path integral, which is a weighted  sum  over all paths. This is not  a ``gauge transformation'' in the traditional sense (since it is not outside of the path integral). This expectation value can be evaluated perturbatively, or more generally, non-perturbatively.

However, it can be viewed as an expectation of the signal model drift line integral operator, i.e., 
\begin{align}\label{eq:WilsonLine}
	\exp\left( \frac{1}{\hbar_{\nu}}\sum_{i=1}^n\int_{x(t_0)=x_0}^{x(t)=x}d\hat{x}_i(t)f_i(\hat{x}(t)) \right),
\end{align}
in the quantum mechanical system with Lagrangian given by Equation \ref{eq:LagEquiv}.

Note that there are several differences between this operator and the Wilson lines that arise in quantum field theories. First of all, this is a quantum mechanical system, not a quantum field theory. Secondly, in field theory the coordinates are a parameter, whereas here (as in quantum mechanics) the coordinates are quantized and not an operator. 


This can be summarized as follows: \textit{The fundamental solution of the Yau Equation, Equation \ref{eq:yauONFeqn}, is the same as the matrix element of the operator in Equation \ref{eq:WilsonLine} for the Euclidean quantum mechanical system with the Hamiltonian as in Equation \ref{eq:EqHamiltonian}.	}

\subsection{A Special Case}

The quantum mechanical equivalence can be made even more direct for a certain class of the signal model drift. This reault is known and shown here using path integral methods. Specifically, suppose the line integral is path independent. Then, it can be factored out of the path integral since its value does not depend on the path. It is straightforward to see when this is possible.

In the one-dimensional case, it is always possible to write the (smooth) drift as a gradient:
\begin{align}
	f(x)=\frac{d}{dx}g(x),
\end{align}
since $g(x)$ is simply given by the integral of $f(x)$. 

In the general $n-$dimensional case, path independence of the line integral  requires that the drift be a gradient:
\begin{align}
	f(x)=\nabla\phi(x).
\end{align}
A special case is the linear symmetric case. Specifically, if $L$ is a symmetric matrix the drift as
\begin{align}
	f(x)&=\sum_{j=1}^n[L_{ij}x_j+l_i],\\ \nonumber
	&=\nabla\phi(x),
\end{align}
where 
\begin{align}
	\phi(x)=\frac{1}{2}\sum_{i,j=1}^nx_iL_{ij}x_j+\sum_{i=1}^nl_ix_i.
\end{align}
Clearly, the the state model drift for the Yau filtering system with $L$ a symmetric matrix is also a gradient of some scalar. 

In all such cases, the expectation of the relevant operator is simply
\begin{align}
	\exp\left( \frac{1}{\hbar_{\nu}}\left[ \phi(x)-\phi(x_0) \right] \right)\int_{x(t_0)=x_0}^{x(t)=x}[\mc{D}x(t)]\exp\left(-\frac{1}{\hbar_{\nu}}\int_{t_0}^tdt\mc{L}\right).
\end{align}
In fact,
\begin{align}\label{eq:RelEQMONF}
	\tilde{P}(t,x|t_0,x_0)=\exp\left( \frac{1}{\hbar_{\nu}}\left[ \phi(x)-\phi(x_0) \right] \right)\vev{x,t|x_0,t_0}.
\end{align}
A gauge transformation relates the fundamental solution of the Yau Equation and the corresponding Euclidean quantum mechanical system.

The discussion in this subsection can be summarized as follows:	\textit{When the signal model drift is a gradient of a scalar field, the fundamental solution of the Yau Equation is, up to a gauge transformation, the same as the transition probability amplitude of a Euclidean quantum mechanical system with Lagrangian given by Equation \ref{eq:LagEquiv}. }

\section{Nonlinear Yau Filtering System and Time-Dependent Schr\"odinger equation}
\subsection{The Work of S-T. Yau and Stephen Yau}

As discussed in Section \ref{sec:BasicResults}, in order to obtain the unnormalized conditional probability $\sigma(t,x)$, it suffices to solve the YYe
\begin{align}
{\left\lbrace\begin{aligned}	%
		\frac{\partial\tilde{u}}{\partial t}(t,x)&=\frac{1}{2}\Delta\tilde{u}(t,x)-\sum_{j=1}^nf_j(x)\frac{\partial\tilde{u}}{\partial x_j}(t,x)-\left( \sum_{j=1}^n\frac{\partial f_j}{\partial x_j}(x)+\frac{1}{2}\sum_{j=1}^mh_j^2(x) \right)\tilde{u}(t,x),\\ 
		\tilde{u}(\tau_{i-1},x)&=\sigma_i(x).\end{aligned}
\right.}
\end{align}

In \cite{YauS-T.Yau2004}, the authors considered the filtering system signal model drift
\begin{align}
	f(x)=Lx+l+\nabla\phi,
\end{align}
where $L$ is an anti-symmetric matrix. The symmetric part can be incorporated in $\phi$. By considering the quantity $\tilde{\nu}(t,x)$ defined by
\begin{align}
	\tilde{u}(t,x)&=e^{\phi(x)}\tilde{\nu}(t,x),\\ \nonumber
	&=e^{\phi(x)}\nu(t,\tilde{x}),\qquad 	\tilde{x}(t)=B(t)x(t)+b(t),
\end{align}
they showed that it is sufficient to solve the following Schr\"odinger equation\begin{align}
	\begin{cases}
		\frac{\partial\nu}{\partial t}(t,\tilde{x})&=\frac{1}{2}\nabla\nu(t,\tilde{x})-\frac{1}{2}q\left( B^{-1}(t)\tilde{x}-B^{-1}(t)b(t) \right)\nu(t,\tilde{x}),\\
		\nu(\tau_{i-1},x)&=\sigma_i(x)e^{-\phi(x)},
	\end{cases}
\end{align}
where
\begin{align}\label{eq:Defq}
	q(x)&\equiv\nabla^2\phi(x)+|\nabla\phi|^2(x)+2(Lx+l)\cdot\nabla\phi+\sum_{i=1}^mh_i^2(x)+\tr L,
\end{align}
and
\begin{align}
	B(t)=e^{-Lt},\qquad b(t)=-\int_0^te^{-Ls}lds.
\end{align}
Here,
\begin{align}
	\frac{dB}{dt}(t)=-B(t)L,\qquad\text{and}\qquad \frac{db}{dt}(t)=-B(t)l,
\end{align}
and $B(t)$ is an orthogonal matrix.

\subsection{Path Integral Derivation}
We now derive using path integral methods the result discussed in the previous sectoion relating the nonlinear Yau filtering system to a Schr\"odinger equation using path integral methods. In matrix notation, the Lagrangian for the Yau filter state model is
\begin{align}
	\mc{L}=\frac{1}{2}\left( \dot{x}-Lx-l-\nabla\phi \right)^2+\frac{\hbar_{\nu}}{2}\nabla\cdot f+\frac{\hbar_{\nu}}{2\hbar_{\mu}}\sum_{k=1}^mh_k^2(x). 
\end{align}
Since
\begin{align}
	\nabla\cdot f=\tr L+\nabla^2\phi,
\end{align}
it follows that 
\begin{align}
	\mc{L}&=\frac{1}{2}(\dot{x}-Lx-l)^2+\frac{1}{2}(\nabla\phi)^2(x)-\dot{x}\cdot\nabla\phi+(Lx+l)\cdot\nabla\phi+\tr L+\nabla^2\phi+\frac{\hbar_{\nu}}{2\hbar_{\mu}}\sum_{k=1}^mh_k^2(x),\\ \nonumber
	&=\frac{1}{2}\left( \dot{x}-Lx-l \right)^2-\dot{x}\cdot\nabla\phi+\frac{1}{2}q(x),
\end{align}
where $q(x)$ is as defined in Equation \ref{eq:Defq}. The $\dot{x}\cdot\nabla\phi(x)$ term can be integrated out of the path integral and simply yields a phase factor
\begin{align}\label{eq:PhaseFactor}
	\exp\left( \frac{1}{\hbar_{\nu}}\left[ \phi(x)-\phi(x_0) \right] \right).
\end{align}
Now, in terms of $\tilde{x}(t)$ 
\begin{align}
	x(t)&=B^{-1}(t)[\tilde{x}(t)-b(t)],\\
	\dot{x}(t)&=B^{-1}(t)[\dot{\tilde{x}}(t)+L\tilde{x}(t)+B(t)l-Lb(t)],\\ \nonumber
	Lx(t)&=B^{-1}(t)[L\tilde{x}(t)-Lb(t)]
\end{align}
so that
\begin{align}
	\dot{x}(t)-Lx(t)-l&=B^{-1}(t)\dot{\tilde{x}}(t),\qquad\text{and }\\
	(\dot{x}-Lx-l)^2&=\dot{\tilde{x}}^2. 
\end{align}
Therefore, in terms of $\tilde{x}(t)$, the Lagrangian is
\begin{align}\label{eq:TildeL}
	\mc{L}=\frac{1}{2}\dot{\tilde{x}}^2+\frac{1}{2}q\left( B^{-1}(t)[\tilde{x}-b(t)] \right).
\end{align}
Combining this with Equation \ref{eq:PhaseFactor}, the fundamental solution becomes
\begin{align}\label{eq:PIderYauYauS}
	\tilde{P}(t,x|t_0,x_0)&=\exp\left( \frac{1}{\hbar_{\nu}}[\phi(x)-\phi(x_0)] \right)\int_{\tilde{x}(t_0)=\tilde{x}_0}^{\tilde{x}(t)=\tilde{x}}\left[ \mc{D}\tilde{x}(t) \right]\exp\left( -\frac{1}{\hbar_{\nu}}\int_{t_0}^tdt\mc{L} \right),\\ \nonumber
	&=\exp\left( \frac{1}{\hbar_{\nu}}[\phi(x)-\phi(x_0)] \right)\tilde{\nu}(t,x|t_0,x_0),
\end{align}
where $\mc{L}$ is given by Equation \ref{eq:TildeL}, and this equation defines $\tilde{\nu}(t,x|t_0,x_0)$. Note that the Jacobian of the transformation from the measure $\left[ \mc{D}x(t) \right]$ to $\left[ \mc{D}\tilde{x}(t) \right]$ is unity since $B(t)$ is an orthogonal matrix. 

It follows from the standard path integral formulation of quantum mechanics that $\tilde{\nu}(t,x|t_0,x_0)$ is just the path integral representation of the fundamental solution of the following Schr\"odinger equation:
\begin{align}
	\frac{\partial\nu}{\partial t}(t,\tilde{x})&=\frac{1}{2}\nabla^2\nu(t,\tilde{x})-\frac{1}{2}q\left( B^{-1}(t)\tilde{x}-B^{-1}(t)b(t) \right)\nu(t,\tilde{x}).
\end{align}
This is precisely the equation obtained in \cite{YauS-T.Yau2004}. For the initial condition, since 
\begin{align}
	\tilde{u}(t,x)	&=\int\tilde{P}(t,x|t_0,x_0)\sigma(t_0,x_0)d^nx_0,\\ \nonumber
	&=\exp\left(\frac{1}{\hbar_{\nu}}\phi(x)\right)\tilde{\nu}(t,x)\\ \nonumber
	&=\exp\left(\frac{1}{\hbar_{\nu}}\phi(x)\right)\int\tilde{\nu}(t,x|t_0,x_0)\exp\left(-\frac{1}{\hbar_{\nu}}\phi(x_0)\right)\sigma(t_0,x_0)d^n\tilde{x}_0,  %
\end{align}
it follows that
\begin{align}
	\tilde{P}(t,x|t_0,x_0)=\exp\left( \frac{1}{\hbar_{\nu}}\left[ \phi(x)-\phi(x_0) \right] \right)\tilde{\nu}(t,x|t_0,x_0), 
\end{align}
which is the same as Equation \ref{eq:PIderYauYauS}. Thus the result in \cite{YauS-T.Yau2004} has been independently established.

\section{Additional Remarks}\label{sec:AdditionalRemarks}

A few comments on our results are in order.

\begin{enumerate}
	\item The general equivalence proposed in Section \ref{ssec:AGenEquiv} is that between the fundamental solution of the FPKfe and a certain matrix element of a quantum mechanical system. Note that all previous equivalences were between fundamental solutions of FPKfe and a Schr\"odinger equation. While the previous equivalences were obtainable using operator methods, it is not clear how the proposed general equivalence can be derived using operator methods. 
	\item The Feynman path integral methods used here are very different from the measure-theoretic methods used to study the nonlinear filtering problem (see, for instance, \cite{PierreDelMoral2004}). The Feynman path integral approach developed in \cite{PAPER2,Balaji2008} and this paper  leads to an independent way of tackling the universal nonlinear filtering problem. In particular, unlike the standard filtering theory approaches, the DMZ equation (or its variants) is not taken as an input. On the contrary, the YYe is obtained directly as a consequence of the path integral approach. It is also noted that the Euclidean quantum physics referred to in the equivalence to nonlinear filtering developed in this paper is a quantum mechanical one, not a quantum field theoretical one. Also note that $\hbar_{\nu}$ here is analogous to the Planck's constant, $\hbar$, in quantum physics. 

	\item Although a formal equivalence has been developed between Euclidean quantum mechanics and universal nonlinear filtering, it is important to point out that there is a profound difference between classical and quantum probabilities. In the ``real time'' (as opposed to Euclidean time) quantum mechanics, the transition probability amplitude is not a probability; it may not even be real. The probability amplitude is to be multiplied with its complex conjugate  to obtain a probability density. In contrast, the transition probability density in filtering theory is a classical probability density. We have shown that, in a mathematical sense, matrix elements in a Euclidean quantum mechanical system equal the transition probability density of a classical stochastic process. This equivalence is purely mathematical, not conceptual.  

	\item The reason for some of the mathematical equivalence between nonlinear filtering and quantum physics is the semi-group property. That is, in stochastic processes the Chapman-Kolmogorov semi-group property is a fundamental property of Markov processes. The Chapman-Kolmogorov semi-group property allows the transition probability density in stochastic process to be written as follows. Let us partition the time interval $[t_0,t]$ into $N$ equi-spaced time intervals so that $t_i=t_0+i\epsilon$ where $\epsilon=(t-t_0)/N$. Then, from the Chapman-Kolmogorov semi-group property it follows that
\begin{align}\label{eq:SplitandCK}
	P(t,x|t_0,x_0)&=\int_{x(t_0)=x_0}^{x(t)=x}\{d^nx(t_1)\cdots d^nx(t_{N-1})\}P(t,x|t_{N-1},x(t_{N-1}))\cdots P(t_1,x(t_1)|t_0,x_0),\\ \nonumber
	&=\int_{x(t_0)=x_0}^{x(t)=x}\left\{ \prod_{i=1}^{N-1}d^nx(t_i) \right\}\prod_{i=1}^{N}P(t_i,x(t_i)|t_{i-1},x(t_{i-1})),
\end{align}
where  $x(t_0)=x_0$ and $x(t_N)=x$ and the delta function condition in the definition of $P(t,x|t_0,x_0)$ is written as integration limits. On the other hand, in quantum physics, the semi-group property is a basic property of the time-evolution operator. It is the semi-group property of the time-evolution operator that leads to the path integral formula for the transition probability amplitude.
	\item Note that the unitarity of the time evolution operator plays a crucial role in quantum mechanics. This is because it is essential for conservation of probability. It is noted that the unitarity of the time evolution operator implies that the Hamiltonian operator is Hermitian\footnote{In the standard formulation of quantum mechanics, Hermiticity of the Hamiltonian is required in order to ensure that the eigenvalues of the Hamiltonian (and hence the possible energies) are real and that the time evolution is unitary (i.e., conserves probability). This can also be ensured even if the Hamiltonian is not Hermitian provided that the Hamiltonian is space-time (or $\mc{PT}$) reflection symmetric.  For a pedagogical discussion of non-Hermitian quantum mechanics, see \cite{C.M.BenderD.C.BrodyB.K.Meister2003}.}. 

The evolution of the probability distribution for a continuous Markov processes is described by the FPKfe. The FPKfe operator is not a Hermitian operator. This is not inconsistent with the conservation of probability. This is because the FPKfe is a continuity equation, and so the probability is conserved (as long as the boundary terms vanish). This is yet another instance of the profound difference between classical and quantum probabilities. 

\item In the continuous-continuous filtering case, the YYe plays the role that the FPKfe plays in continuous-discrete filtering. However, the YYe is not a continuity equation. This is not a contradiction since the YYe is related to the unnormalized conditional probability density, whereas the FPKfe evolves (and preserves the normalization of) the normalized probability density. 

\item In this paper, it has been assumed that the noise is additive and the model is not explicitly time-dependent. In the more general case, a simple  equivalence, as derived in this paper, is not possible. For instance, in order to obtain the YYe, it was necessary to assume that the measurement model was not explicitly time dependent (see \cite{PAPER2}); this is not valid in the general case. Also, when the noise is multiplicative, quantization is not as straightforward. This can be traced to the well-known  operator ordering ambiguity in quantum physics (since the position and momentum operators do not commute). Furthermore, standard calculus manipulations are no longer possible in the path integral. However, the path integral result is the same even for the multiplicative noise case if the diffusion matrix is proportional to the identity matrix.

\end{enumerate}

Finally, note that the results of this paper also apply to the FPKfe itself, which is simply the case $h(x)=0$, in a straightforward manner.

\section{Conclusions and Future Work}\label{sec:Conclusion}

The main conclusion of this paper is that certain continous-continuous nonlinear filtering problems are related to Euclidean quantum mechanics. Specifically, the transition probability density for nonlinear filtering problems  with additive noise and with square diffusion vielbein and explicitly time-independent drift is the fundamental solution of the YYe, and is the same as the expectation of a certain operator in a quantum mechanical system. A corollary of this general result is a derivation of the relationship between the YYe in the Yau filtering system and a time-varying Schr\"odinger equation which was first derived earlier. This equivalence leads to some useful results from the methods used in quantum physics.  

There are many possible directions for future work:
\begin{enumerate}
	\item The path integral formula can be exactly solvable, or lead to simpler methods of obtaining solutions, in some simple cases. In fact, for the model studied  in \cite{YauS-T.Yau2004}, an explicit path integral solution is described in a paper currently in preparation \cite{EXSOLYAU1}.
	\item The equivalence to Euclidean quantum mechanics immediately leads to many filtering problems that can be solved exactly, as will be explored in future papers. The remarkable fact is that many exactly solvable Euclidean quantum mechanical problems correspond to filtering problems that are not finite-dimensional (i.e., Lie algebra of $\mc{L}_0$ and $h_i(x), i=1,\dots,m$ is not finite-dimensional, see \cite{PAPER2}) . Thus, \textit{simplicity in filtering theory does not imply finite dimensionality}.   
		
	\item The path integral formulation naturally leads to a perturbative solution of the nonlinear filtering problem. Such a perturbative solution which is analogous to extended Kalman filtering is called extended Yau filtering (EYF). Thus, it is possible to perturb about the Yau filter (a generalization of the linear Kalman filter), rather than the linear Kalman filter. Clearly, since the EKF is a special case of the EYF, such a formulation will be superior to  the EKF. 
	\item However, it is noted that the fundamental solution is defined nonperturbatively. This is important since sometimes the perturbative approaches, like EKF, fail. 
	\item Perhaps the most important advantage of the path integral lies in numerical methods of computation \cite{IstvanMontvayandGernotMunster1997}. In fact, path integrals are the only known way to carry out nonperturbative computation in quantum chromodynamics (QCD). Note that QCD, the gauge theory of strong interactions, is a quantum field theory, not merely quantum mechanical and the QCD Lagrangian is highly nonlinear. The excellent numerical results in QCD suggest that currently known path integral methods  should give very good performance for the simpler case of (large dimensional) universal nonlinear filtering. It is sufficient to note that the Dirac-Feynman approximation, the crudest approximation of the path integral, already yields excellent results (see \cite{Balaji2008}), and is adequate for smaller dimensional problems.  

\end{enumerate}

It is clear that many lines of investigation are suggested by the path integral methods and it is planned that results of those investigations will be presented in future papers. 

\appendix
\section{Verification of the Path Integral Formula}\label{sec:VerifyYau}

It is now proven that the path integral formula Equation \ref{eq:PITimeIndepSamp} satisfies the YYe. This closely follows the method used by Feynman to verify the path integral formula for the Schr\"odinger equation in quantum mechanics.

According to the Chapman-Kolmogorov semi-group property 
\begin{align}
	\tilde{P}(t+\epsilon,x|t_0,x_0)=\int \tilde{P}(t+\epsilon,x|t,x')\tilde{P}(t,x'|t_0,x_0)\{d^nx'\},
\end{align}
where $\tilde{P}(t,x'|t_0,x_0)$ is given by Equation \ref{eq:Yaukernel}. %
When $\epsilon$ is an infinitesimal
\begin{align}\label{eq:CKVerify}
	\tilde{P}&(t+\epsilon,x|t,x')=\\ \nonumber
	&\exp\left(-\frac{1}{2\hbar_{\nu}\epsilon}\sum_{i=1}^n\left[ x_i-x_i'-\epsilon f_i(\bar{x}) \right]^2-\epsilon\frac{1}{2}\sum_{i=1}^n\frac{\partial f_i}{\partial x_i}(\bar{x})-\epsilon\frac{1}{2\hbar_{\mu}}\sum_{k=1}^mh_k^2(\bar{x})\right),
\end{align}
where $\bar{x}=\frac{1}{2}(x+x')$. The dominant contribution occurs when the following condition is satisfied:
\begin{align}
	x-x'-\epsilon f(\bar{x})\approx0.
\end{align}
We may write in this region
\begin{align}
	x&=x'+\epsilon f(\bar{x})+\eta,\qquad\text{or}\\ \nonumber
	x&=x'+\epsilon f(x)+\eta,
\end{align}
where the equalities are valid to $O(\epsilon)$. Substituting this into Equation \ref{eq:CKVerify} (keeping  terms up to $O(\epsilon)$)
\begin{align}\label{eq:PathIntegFormVer} 
	\tilde{P}(t+\epsilon,x|t_0,x_0)&=A\int_{-\infty}^{\infty}\exp\left(-\frac{1}{2\hbar_{\nu}\epsilon}\sum_{i=1}^n\eta_i^2\right) \\ \nonumber
	& \quad\times\left( 1-\frac{\epsilon}{2}\sum_{i=1}^n\frac{\partial f_i}{\partial x_i}(x) -\frac{\epsilon}{2\hbar_{\mu}}\sum_{k=1}^mh_k^2(x)\right)\tilde{P}(t,x'|t_0,x_0)\left\{ d^nx' \right\},\\ \nonumber
	&=A\int_{-\infty}^{\infty}\exp\left(-\frac{1}{2\hbar_{\nu}\epsilon}\sum_{i=1}^n\eta_i^2\right)& \\ \nonumber
& \quad\times\left( 1-\frac{\epsilon}{2}\sum_{i=1}^n\frac{\partial f_i}{\partial x_i}(x) -\frac{\epsilon}{2\hbar_{\mu}}\sum_{k=1}^mh_k^2(x)\right)\tilde{P}(t,x'|t_0,x_0)\left( 1-\frac{\epsilon}{2}\sum_{i=1}^n\frac{\partial f_i}{\partial x_i}(x) \right)\left\{d^n\eta\right\}, \\ \nonumber
&=A\int_{-\infty}^{\infty}\left\{ d^n\eta \right\}\exp\left(-\frac{1}{2\hbar_{\nu}\epsilon}\sum_{i=1}^n\eta_i^2\right)\left( 1-\epsilon\sum_{i=1}^n\frac{\partial f_i}{\partial x_i}(x) -\frac{\epsilon}{2\hbar_{\mu}}\sum_{k=1}^mh_k^2(x)\right)\\ \nonumber
&\qquad \tilde{P}(t,x-\epsilon f(x,t)-\eta|t_0,x_0).%
\end{align}
It is noted that the Jacobian of the the transformation from $x_0$ to $\eta$ to $O(\epsilon)$ is included in the second step.

The constant $A$ is fixed by 
\begin{align}
\label{eq:Anormaliz}
A\int_{-\infty}^{\infty}\exp\left( -\frac{1}{2\hbar_{\nu}\epsilon}\sum_{i=1}^n\eta_{i}^2 \right)\left\{ d^n\eta \right\}=1.
\end{align}
Hence, it follows that
\begin{align}
	A\int_{-\infty}^{\infty}\eta_i\eta_j\exp\left( -\frac{1}{2\hbar_{\nu}\epsilon}\sum_{i=1}^n\eta_i^2 \right)\left\{ d^n\eta \right\}=\hbar_{\nu}\epsilon\delta_{ij}.
\end{align}

The left hand side of Equation \ref{eq:PathIntegFormVer} is 
\begin{align}
	\tilde{P}(t,x|t_0,x_0)+\epsilon\frac{\partial \tilde{P}}{\partial t}(t,x|t_0,x_0).
\end{align}
The second order expansion (in $\eta$) of  $P(t,x'|t_0,x_0)$ yields  
\begin{align}
	\tilde{P}(t,x'|t_0,x_0)&=\tilde{P}(t,x-\epsilon f(x,t)-\eta|t_0,x_0),\\ \nonumber
	&=\tilde{P}(t,x|t_0,x_0)\left(1-\epsilon\sum_{i=1}^n\frac{\partial f_i}{\partial x_i}(x,t)-\epsilon\frac{1}{2\hbar_{\mu}}\sum_{k=1}^mh_k^2(x)\right)\\ \nonumber
	&\qquad-\sum_{i=1}^n\left( \epsilon f_i(x)+\eta_i \right)\frac{\partial \tilde{P}}{\partial x_i}(t
	,x|t_0,x_0)+\frac{1}{2}\sum_{i,j=1}^n\eta_i\eta_j\frac{\partial^2\tilde{P}}{\partial x_i\partial x_j}(t,x|t_0,x_0).
\end{align}
We are interested only in terms of $O(\epsilon).$  The only nonvanishing terms are terms linear in $\epsilon$ and  quadratic in $\eta$. Then it is easy to verify that the diffusion part  of the YYe follows from the term quadratic in $\eta$ and the rest follows from the term linear in $\epsilon$, so that 
\begin{align}
{\left\lbrace\begin{aligned}
\frac{\partial \tilde{P}}{\partial t}(t,x|t_0,x_0)&=\frac{\hbar_{\nu}}{2}\sum_{i=1}^n\frac{\partial^2\tilde{P}}{\partial x_i^2}(t,x|t_0,x_0)-\sum_{i=1}^n\frac{\partial}{\partial x_i}\left[ f_i(x)\tilde{P}(t,x|t_0,x_0) \right]%
-\frac{1}{2\hbar_{\mu}}\sum_{k=1}^mh_k^2(x)\tilde{P}(t,x|t_0,x_0),\\ %
	\tilde{P}(t_0,x|t_0,x_0)&=\delta^n(x-x_0).\end{aligned}
\right.}
\end{align}
Hence, $\tilde{P}$ satisfies the YYe with a delta-function initial condition. %

\bibliographystyle{utphys}
\bibliography{onfbib}

\providecommand{\href}[2]{#2}\begingroup\raggedright\begin{thebibliography}{10}

\bibitem{H.Risken1999}
H.~Risken, {\em The {Fokker-Planck} Equation: Methods of Solution and
  applications}.
\newblock Springer-Verlag, second~ed., 1999.

\bibitem{Balaji2008}
B.~Balaji, ``Estimation of indirectly observable langevin states: path integral
  solution using statistical physics methods,''
  \href{http://dx.doi.org/10.1088/1742-5468/2008/01/P01014}{{\em Journal of
  Statistical Mechanics: Theory and Experiment} {\bf 2008} (2008) no.~01,
  P01014 (17pp)}.

\bibitem{A.H.Jazwinski1970}
A.~H. Jazwinski, {\em Stochastic Processes and Filtering Theory}.
\newblock Dover Publications, 2007.

\bibitem{PAPER2}
B.~Balaji, ``Universal nonlinear filtering using path integrals {II}: The
  continuous-continuous model with additive noise,''
  \href{http://arxiv.org/abs/arXiv:0708.1663}{{\tt arXiv:0708.1663}}.

\bibitem{S.T.YauS.-TYau2000}
S.~T. Yau and S.~S.-T. Yau, ``Real time solution of nonlinear filtering problem
  without memory {I},'' {\em Mathematical Research Letters} {\bf 7} (2000)
  671--693. \url{www.mrlonline.org/mrl/2000-007-006/2000-007-006-002.pdf}.

\bibitem{R.P.FeynmanandA.R.Hibbs1965}
R.~P. Feynman and A.~R. Hibbs, {\em Quantum Mechanics and Path Integrals}.
\newblock McGraw-Hill Book Company, 1965.

\bibitem{YauS-T.Yau2004}
S.-T. Yau and S.~S.-T. Yau, ``Nonlinear filtering and time varying
  schr\"odinger equation {I},''{\em IEEE Transactions on Aerospace and
  Electronic Systems} {\bf 40} (January, 2004)  284--292.

\bibitem{S.-T.YauS.S.-TYau1996}
S.-T. Yau and S.~S.-T. Yau, ``Explicit solution of a kolmogorov equation,''
  \href{http://dx.doi.org/10.1007/s002459900028}{{\em Applied Mathematics and
  Optimization} {\bf 34} (1996) no.~3, 231--266}.

\bibitem{S.S.-T.YauandS.-T.Yau2001}
S.~S.-T. Yau and S.-T. Yau, ``Real time algorithm for nonlinear filtering
  problem,'' in {\em Proceedings of the 40th IEEE Conference on Decision and
  Control}.
\newblock Orlando, FL, USA, December, 2001.

\bibitem{YauYau2005}
S.~S.-T. Yau and S.-T. Yau, ``Solution of filtering problem with nonlinear
  observations,'' {\em SIAM Journal of Control and Optimization} {\bf 44}
  (2005) no.~3, 1019--1039.

\bibitem{YauYau2008}
S.-T. Yau and S.~S.-T. Yau, ``Real time solution of the nonlinear filtering
  problem without memory {II},''
  \href{http://dx.doi.org/10.1137/050648353}{{\em SIAM Journal on Control and
  Optimization} {\bf 47} (2008) no.~1, 163--195}.
  \url{http://link.aip.org/link/?SJC/47/163/1}.

\bibitem{PAPER1}
B.~Balaji, ``Universal nonlinear filtering using path integrals {I}: The
  continuous-discrete model with additive noise,'' {\em submitted to IEEE
  Transactions on Aerospace and Electronic Systems} (2006)  ,
  \href{http://arxiv.org/abs/arXiv:0708.0354}{{\tt arXiv:0708.0354}}.
  \url{http://www.citebase.org/abstract?id=oai:arXiv.org:0708.0354}.

\bibitem{JeanZinn-Justin2002}
J.~Zinn-Justin, {\em Quantum Field Theory and Critical Phenomena}.
\newblock Int. Ser. Monogr. Phys. Oxford University Press, 2002.

\bibitem{PierreDelMoral2004}
P.~D. Moral, {\em Feynman-K\v ac Formulae}.
\newblock Springer-Verlag, March, 2004.

\bibitem{C.M.BenderD.C.BrodyB.K.Meister2003}
C.~M. Bender, D.~C. Brody, and B.~K. Meister, ``Must a hamiltonian be
  hermitian?,'' {\em American Journal of Physics} {\bf 71} (2003)  1095--1102.

\bibitem{EXSOLYAU1}
B.~Balaji, ``Exactly solvable nonlinear filtering {I}: The yau filter with
  quadratic $\eta$ and the harmonic oscillator,'' {\em in preparation}  .

\bibitem{IstvanMontvayandGernotMunster1997}
I.~Montv\'ay and G.~M\"unster, {\em Quantum Fields on a Lattice}.
\newblock Cambridge Monographs on Mathematical Physics. Cambridge University
  Press, New York, USA, 1997.

\end{thebibliography}\endgroup

\end{document}